# QUANTUM PHYSICS IN NEUROSCIENCE AND PSYCHOLOGY: A NEW MODEL WITH RESPECT TO MIND/BRAIN INTERACTION

Jeffrey M. Schwartz [1]

Henry P. Stapp [2]

Mario Beauregard [3, 4, 5, 6*]

1 UCLA Neuropsychiatric Institute, 760 Westwood Plaza, C8-619 NPI Los Angeles, California 90024-1759, USA. E-mail: jmschwar@ucla.edu

2 Theoretical Physics Mailstop 5104/50A Lawrence Berkeley National Laboratory, University of California, Berkeley, California 94720-8162, USA. Email: hpstapp@lbl.gov

3 Département de psychologie, Université de Montréal, C.P. 6128, succursale Centre-Ville, Montréal, Québec, Canada, H3C 3J7.

4 Département de radiologie, Université de Montréal, C.P. 6128, succursale Centre-Ville, Montréal, Québec, Canada, H3C 3J7.

5 Centre de recherche en sciences neurologiques (CRSN), Université de Montréal, C.P. 6128, succursale Centre-Ville, Montréal, Québec, Canada, H3C 3J7.

6 Groupe de Recherche en Neuropsychologie Expérimentale et Cognition (GRENEC), Université de Montréal, C.P. 6128, succursale Centre-Ville, Montréal, Québec, Canada, H3C 3J7.

————

*Correspondence should be addressed to: Mario Beauregard, Département de psychologie, Université de Montréal, C.P. 6128, succursale Centre-Ville, Montréal, Québec, Canada, H3C 3J7. Tel (514) 340-3540 #4129; Fax: (514) 340-3548; E-mail: mario.beauregard@umontreal.ca

## Short abstract

Neuropsychological research on the neural basis of behavior generally posits that brain mechanisms fully suffice to explain all psychologically described phenomena. Terms having intrinsic experiential content (e.g., "feeling," "knowing" and "effort") are not included as causal factors because they are deemed superfluous to the causal mechanisms of brain function. However, principles of quantum physics causally relate mental and physical properties. Use of this causal connection allows neuroscientists and psychologists to more adequately and effectively investigate the neuroplastic mechanisms relevant to the growing number of studies of the capacity of directed attention and mental effort to systematically alter brain function.

## Long abstract

The cognitive frame in which most neuropsychological research on the neural basis of behavior is conducted contains the assumption that brain mechanisms *per se* fully suffice to explain all psychologically described phenomena. This assumption stems from the idea that the brain is made up entirely of material particles and fields, and that all causal mechanisms relevant to neuroscience must therefore be formulated solely in terms of properties of these elements. One consequence of this stance is that psychological terms having intrinsic mentalistic and/or experiential content (terms such as "feeling," "knowing" and "effort") have not been included as primary causal factors in neuropsychological research: insofar as properties are not described in material terms they are deemed irrelevant to the causal mechanisms underlying brain function. However, the origin of this demand that experiential realities be excluded from the causal base is a theory of nature that has been known to be fundamentally incorrect for more than three quarters of a century. It is explained here why it is consequently scientifically unwarranted to assume that material factors alone can in principle explain all causal mechanisms relevant to neuroscience. More importantly, it is explained how a key quantum effect can be introduced into brain dynamics in a simple and practical way that provides a rationally coherent, causally formulated, physics-based way of understanding and using the psychological and physical data derived from the growing set of studies of the capacity of directed attention and mental effort to systematically alter brain function.

"[T]he only acceptable point of view appears to be the one that recognizes *both* sides of reality --- the quantitative and the qualitative, the physical and the psychical --- as compatible with each other, and can embrace them simultaneously."

Wolfgang Pauli, *The Influence of Archetypal Ideas on the Scientific Theories of Kepler*

1. Introduction

The introduction into neuroscience and neuropsychology of the extensive use of functional brain imaging technology has led to a major conceptual advance pertaining to the role of directed attention in cerebral functioning. On the empirical side the identification of brain areas involved in a wide variety of information processing functions concerning learning, memory and various kinds of symbol manipulation has been the object of a large amount of intensive investigation (See Toga & Mazziotta 2000). As a result neuroscientists now have a reasonably good working knowledge of the role of a variety of brain areas in the processing of complex information. But, valuable as these empirical studies are, they provide only the data for, not the answer to, the critical question of the causal relationship between the psychologically described information and the central nervous system (CNS) mechanisms that process this information. In the vast majority of cases investigators simply assume that measurable properties of the brain are the only factors needed to explain, at least in principle, all of the types of information processing that are experimentally observed. This privileging of physically describable brain mechanisms as the core, and indeed final, explanatory vehicle for the processing of every kind of psychologically formulated data is, in fact, the foundational assumption of almost all contemporary biologically based cognitive neuroscience.

It is becoming increasingly clear, however, that there is at least one type of information processing and manipulation that does not readily lend itself to explanations that assume that all final causes are subsumed within brain, or more generally, CNS mechanisms. The cases in question are those in which the conscious act of willfully altering the mode by which experiential information is processed itself changes, in systematic ways, the cerebral mechanisms utilized. There is a growing recognition of the theoretical importance of applying experimental paradigms that employ directed mental effort in order to produce systematic and predictable changes in brain function (e.g., Beauregard et al. 2001; Ochsner et al. 2002). These wilfully induced brain changes are generally accomplished through training in the cognitive reattribution and attentional recontextualization of conscious experience. Further, an accelerating number of studies in the neuroimaging literature significantly support the thesis that, again, with appropriate training and effort, people can systematically alter neural circuitry associated with a variety of mental and physical states that are frankly pathological (Schwartz et al. 1996; Schwartz 1998; Musso et al. 1999; Paquette et al. 2003). A recent review of this and the related neurological literature has coined the term "self-directed neuroplasticity" to serve as a general description of the principle that focused training and effort can systematically alter cerebral function in a predictable and potentially therapeutic manner (Schwartz & Begley 2002).

From a theoretical perspective perhaps the most important aspect of this line of empirical research is its direct relevance to new developments in our understanding of the physics of the interface between mind/consciousness and brain. Until recently virtually all attempts to understand the functional activity of the brain have been based ultimately on principles of classical physics that have been known to be fundamentally false for three quarters of a century. A basic feature of that classical conception of the world is that all causal connections are carried by, and are completely explainable in terms of, direct interactions between material realities. This truncated view of causation is not entailed by the current principles of physics, which provide a far more adequate and useful foundation for the description and understanding of the causal structure of self-directed neuroplasticity. The superiority of contemporary physics in this context stems from two basic facts. First, terms such as "feeling," "knowing" and "effort," because they are intrinsically mentalistic and experiential, cannot be described exclusively in terms of material structure. And, second, mentalistic terminology of precisely this kind is critically necessary for the design and execution of the experiments in which the data demonstrating the core phenomena of self-directed neuroplasticity are acquired and described. Thus the strictly materialistic principles of causation to which one is restricted by the form of classical physics enforce a causal and semantic gap between the neurological and psychological parts of the data of self-directed neuroplastic phenomena. On the other hand, physics, as it is currently practiced, utilizes quantum principles that, as we shall explain in detail, fully allow for the scientific integration of mentalistic and neurophysiological terminology. These principles provide for logically coherent rational explanations that are entirely capable of accounting for the causal mechanisms necessary to understand the rapidly emerging field of self-directed neuroplasticity.

In order to explicate the physics of the interface between mind/consciousness and brain, we shall in this article describe in detail just how the quantum mechanically based causal mechanisms work, and show why it is necessary in principle to advance to the quantum level to achieve an adequate understanding of neurophysiology during volitionally directed activity. The reason, basically, is that classical physics is an approximation to the more accurate quantum theory, and this approximation eliminates the causal efficacy of our conscious efforts that is manifested in these experiments. In addition, structural features of ion conductance channels critical to synaptic function require that quantum reasoning must be applied in principle.

The theoretically important point is that classical physics, and the associated doctrine of materialism, fail to coherently explain self-directed neuroplastic phenomena, while the quantum mechanical principles that causally integrate mentalistic and physicalistic data clearly and explicitly do. Because experientially based language is not logically reducible to classical materialist terminology, yet such mentalistic language is a logical pre-requisite for the design, execution, and description of volitionally directed neuroplastic phenomena, the attempt to explain such phenomena in solely materialist terms must be abandoned as a matter of principle: the logical structure of materialism is inadequate in these cases. In the light of the causal structure of quantum physics, as described in some detail in later sections of this article, the case for giving brain mechanisms a privileged position as the sole cause of our conscious efforts, and of their consequences, has become radically atheoretical and ungrounded in reason.

Let us be entirely clear about the sort of neuroscientific reasoning that remains coherent, given the structure of modern physics, and, contrastingly, the types of assertions that should now be viewed as merely the residue and cultural baggage of a materialistic bias stemming from superceded physical concepts. Entirely acceptable are **correlational analyses** concerning the relationship between mentalistic data and neurophysiological mechanisms. Examining the qualitative and quantitative aspects of brain function, and doing detailed analyses of how they relate to the data of experience, obtained through increasingly sophisticated means of psychological investigation and subject self-report analysis (e.g., the entire Sep/Oct 2003 issue of *Journal of Consciousness Studies*, Volume 10, Number 9-10, is dedicated to these questions), can now be seen as being both completely in line with fundamental physics, and also the core structure of neuropsychological science. To a significant degree this is already the case. However, what is not justified is the assumption that *all* aspects of experience examined and reported are necessarily causal consequences solely of brain mechanisms that are in principle observable. The structure of contemporary physics entails no such conclusion. This is particularly relevant to data from first person reports concerning active willfully directed attentional focus, and especially to data regarding which aspects of the stream of conscious awareness a subject chooses to focus on when making self-directed efforts to modify and/or modulate the quality and beam of attention. In such cases the structure of orthodox quantum physics implies that the investigator is not justified in assuming that the focus of attention is determined wholly by brain mechanisms that are in principle completely well defined and mechanically determined. Conscious effort itself can justifiably be taken to be a primary variable whose complete causal origins may be untraceable in principle, but whose causal efficacy in the physical world is real.

The quantum mechanical principles that causally integrate mental and physical phenomena, which are separately taken to be both indispensable and irreducible, provide a rationally coherent foundation for modern neuroscience and neuropsychology.

## 2. Practical and theoretical aspects of self-directed neuroplasticity

The cognitive frame in which neuroscience research, including research on cerebral aspects of behavior, is generally conducted contains within it the assumption that brain mechanisms *per se*, once discovered, are fully sufficient to explain whatever phenomenon is being investigated. In the fields of functional neuroimaging this has led to experimental paradigms that focus primarily on changes in brain tissue activation as primary variables used to explain whatever behavioral changes are observed --- including ones understood as involving essentially cognitive and emotional responses. As long as one is investigating phenomena that are mostly passive in nature this may well be fully justified. A person is shown a picture depicting an emotionally or perhaps a sexually arousing scene. The relevant limbic and/or diencephalic structures are activated. The investigator generally concludes that the observed brain activation has some intrinsic causal role in the emotional changes reported (or perhaps, the hormonal correlates of those changes). All is well and good, as far as it goes. And all quite passive from the experimental subject's perspective --- all that's really required on his or her part is to remain reasonably awake and alert, or, more precisely, at least somewhat responsive to sensory inputs. But when, as happens in a growing number of studies, the subject makes an active response aimed at systematically *altering* the nature of the emotional reaction --- for example, by actively performing a cognitive reattribution --- understanding the data solely from the perspective of brain-based causal mechanism can be severely limiting and counterproductive. This is especially so when one is investigating how to develop improved methods for altering the emotional and cerebral responses to significantly stressful external or internally generated stimuli.

Simply stated, the prevailing prejudices, unsupported by contemporary physics, about the respective causal roles of neurophysiologically and mentalistically described variables seriously limits the scope and utility of the present matter-based theory of conscious-brain activity. While one may immediately grant that that these two types of variables are quite intimately related, and that complete clarity concerning their respective role in any given human action can be difficult (and sometimes even impossible), the fact remains that the serious investigator of human neuropsychology must make a concerted effort to sort out the differences. This is especially so when the phenomena under investigation are value-laden, i.e., involve the possibility of making choices and decisions about how to respond to sensory phenomena.

In the case of studying clinical phenomena such as psychological treatments and their biological effects, the distinction between mind and brain (or, if one prefers, mentalistic and neurophysiological variables) becomes absolutely critical. That's because if one simply assumes the most common generic belief of our era of neuroscience research, namely that all aspects of emotional response are passively determined by neurobiological mechanisms, then the possibility of developing genuinely effective self-directed psychological strategies that cause real neurobiological changes becomes, in principle, impossible. The clinician thus becomes locked, as it were, into at least the implicit view that the psychological treatment of ailments caused by neurobiological impairments is not a realistic goal.

There is already a wealth of data arguing against this view. For instance, work in the 1990's on patients with obsessive compulsive disorder demonstrated significant changes in caudate nucleus metabolism and the functional relationships of the orbitofrontal cortex-striatum-thalamus circuitry in patients who responded to a psychological treatment utilizing cognitive reframing and attentional refocusing as key aspects of the therapeutic intervention (for review see Schwartz & Begley 2002). More recently work by Beauregard and colleagues (Paquette et al. 2003) have demonstrated systematic changes in the dorsolateral prefrontal cortex and parahippocampal gyrus after cognitive-behavioral therapy for spider phobia, with brain changes significantly related to both objective measurements and subjective reports of fear and aversion. There are now numerous reports on the effects of self-directed regulation of emotional response, via cognitive reframing and attentional recontextualization mechanisms, on cerebral function (e.g., Beauregard et al. 2001; Lévesque et al. 2003; Ochsner et al. 2002; Paquette et al. 2003; Schwartz et al. 1996).

The brain area generally activated in all the studies done so far on the self-directed regulation of emotional response is the prefrontal cortex, an area of the brain also activated in studies of cerebral correlates of willful mental activity, particularly those investigating self-initiated action and the act of attending to one's own actions (Spence & Frith 1999; Schwartz & Begley 2002). There is however one aspect of willful mental activity that seems particularly critical to emotional self-regulation and seems to be the critical factor in it's effective application --- the factor of focused dispassionate self-observation that, in a rapidly growing number of clinical psychology studies, has come to be called mindfulness or mindful awareness (Segal et al. 2002)

The mental act of clear-minded introspection and observation, variously known as mindfulness, mindful awareness, bare attention, the impartial spectator, etc. is a well-described psychological phenomenon with a long and distinguished history in the description of human mental states (Nyanaponika 2000). The most systematic and extensive exposition is in the canonical texts of classical Buddhism preserved in the Pali language, a dialect of Sanskrit. Because of the critical importance of this type of close attentiveness in the practice of Buddhist meditation, some of it's most refined descriptions in English are in texts concerned with meditative practice (although it is of critical importance to realize that the mindful mental state does not require any specific meditative practice to acquire, and is *certainly not* in any sense a "trance-like" state). One particularly well-established description, using the name bare attention, is as follows:

> "Bare Attention is the clear and single-minded awareness of what actually happens *to* us and *in* us at the successive moments of perception. It is called 'Bare' because it attends just to the bare facts of a perception as presented either through the five physical senses or through the mind . . . without reacting to them." (Nyanaponika 1973, p.30)

Perhaps the essential characteristic of mindful observation is that you are just watching, observing all facts, both inner and outer, very calmly, clearly, and closely. To sustain this attentional perspective over time, especially during stressful events, invariably requires the conscious application of effort.

A working hypothesis for ongoing investigation in human neurophysiology, based on a significant body of preliminary data, is that the mental action of mindful awareness specifically modulates the activity of the prefrontal cortex. Because of the well established role of this cortical area in the planning and willful selection of self-initiated responses (Spence & Frith 1999; Schwartz & Begley 2002), the capacity of mindful awareness, and by implication all emotional self-regulating strategies, to specifically modulate activity in this critical brain region has tremendous implications for the fields of mental health and related areas.

The major theoretical issue we are attempting to address in this article is the failure of classical models of neurobiological action to provide a scientifically adequate account for all of the mechanisms that are operating when human beings utilize self-directed strategies for the purpose of modulating emotional responses and their cerebral correlates. Specifically, the assumption that all aspects of mental activity and emotional life are ultimately explicable solely in terms of micro-local deterministic brain activity, with no superposed effects of mental effort, produces a theoretical structure that both fails to meet practical scientific needs, and also fails to accommodate the causal structure of modern physics. The simple classical model must in principle be replaced by the physically more accurate and functionally more useful concept in which the role played by the mind, when observing and modulating one's own emotional states, is an intrinsically active and physically efficacious process in which mental action is *affecting* brain activity, not merely being affected by it. One key reason for the necessity of this change in perspective is the fact that recognition of the active character of the mind in emotional self-regulation is needed both to subjectively access the phenomena (e.g., effort is required to sustain mindfulness during stressful events), and to objectively describe what is subjectively happening when a person directs his or her inner resources to the challenging task of modifying emotional and cerebral responses. It takes *effort* for people to achieve these results. That is because it requires a redirection of the brain's resources away from lower level limbic responses and toward higher level prefrontal functions --- and this does not happen passively. Rather, it requires willful training and directed effort. It is semantically inconsistent and clinically counter productive to insist that these kinds of brain changes be viewed as being solely an intra-cerebral "the physical brain changing itself" type of action. That is because features of the activity essential to its identification, activation, and use are not describable solely in terms of material brain mechanisms.

Furthermore, as we will see in detail in the following sections of this article, orthodox concepts of contemporary physics are ideally suited to a rational and practically useful understanding of the action of mindful self-observation on brain function. Classical models of physics, which view all action in the physical world as being ultimately the result of the movements of material particles, are now seriously out of date, and no longer should be seen as providing the only, or the best, paradigm for investigating the interface between mind/consciousness and brain.

Does it make scientific good sense to try to understand the process of self-directed neuroplasticity solely in terms of brain mechanisms?

For at least one quite straightforward reason it seems clear that it does not. That reason is that it is intrinsically impossible to explain and describe to real people the techniques they must learn to perform and strategies required to initiate and sustain self-directed neuroplastic changes without using language that contains instructions about what to do with your mind, i.e., without using terms referring to mental experience, words like: feeling, effort, observation, awareness, mindfulness, and so forth. When people practice self-directed activities for the purpose of systematically altering patterns of cerebral activation they are attending to their mental and emotional *experiences*, not merely their limbic or hypothalamic brain mechanisms. And while no scientifically oriented person denies that those brain mechanisms play a critical role in generating those experiences, precisely what the person is training himself to do is to willfully *change* how those brain mechanisms operate --- and to do that absolutely requires attending to mental experience *per se*. It is in fact the basic thesis of self-directed neuroplasticity research that the *way* in which a person directs his attention, e.g., mindfully or unmindfully, will affect both the experiential state of the person and the state of his/her brain.

The very acquisition of the skills required in order to change the brain, especially in the attempt to alleviate stressful and/or patholological conditions, requires understanding what it means to observe mindfully etc., and learning those skills cannot be accomplished via the sole use of neurobiological terminology --- the language of mental experience must of necessity be utilized. A growing body of research informs us that when people learn to systematically alter their emotional and/or behavioral responses to stressful stimuli it modulates the activity of the prefrontal cortex, among other areas. But to merely say to someone "Now modulate your prefrontal cortex," just like that, is not, in and of itself, a meaningful use of language. This is so because in the absence of some kind of learning and/or training process that in principle *must* make use of the language of personal experience, it is intrinsically impossible for any real living person to know *how to* modulate their prefrontal cortex. For experimental subjects to actually learn and operationalize the skills and techniques necessary for the collection of the data that demonstrate the phenomena of self-directed neuroplasticity the use of mind-based experiential language is *required.* The assertion that a science of self-directed action can be pursued within a purely materialist framework is neither semantically coherent, nor empirically established, nor rationally entailed by the principles of modern physics.

People can certainly learn how to be mindful, and when they do it, they change brain function in very beneficial ways. But to effect and accomplish those brain changes requires the use of the language of mental experience and activity in basic and irreducible ways --- it can never be accomplished solely by the use of brain-based language. This straightforward fact tells us that the language of neurobiology will never be sufficient for the effective self-regulation of brain activity. The language of the active mind is an absolute logical requirement. We will now see that contemporary physical theory contains a prepared place for the needed causal intervention in brain activity of conscious volition.

3. Classical physics

Classical physics is a theory of nature that originated with the work of Isaac Newton in the seventeenth century and was advanced by the contributions of James Clerk Maxwell and Albert Einstein. Newton based his theory on the work of Johannes Kepler, who found that the planets appeared to move in accordance with a simple mathematical law, and in ways wholly determined by their spatial relationships to other objects. Those motions were apparently *independent of our human observations of them*.

Newton effectively assumed that all physical objects were made of tiny miniaturized versions of the planets, which, like the planets, moved in accordance with simple mathematical laws, independently of whether we observed them of not. He found that he could then explain the motions of the planets, and also the motions of large terrestrial objects and systems, such as cannon balls, falling apples, and the tides, by assuming that every tiny planet-like particle in the solar system attracted every other one with a force inversely proportional the square of the distance between them.

This force was an *instantaneous action at a distance*: it acted instantaneously, no matter how far the particles were apart. This feature troubled Newton. He wrote to a friend "That one body should act upon another through the vacuum, without the mediation of anything else, by and through which their action and force may be conveyed from one to another, is to me so great an absurdity that I believe no man, who has in philosophical matters a competent faculty of thinking, can ever fall into it." (Newton 1687: 634) Although Newton's philosophical persuasion on this point is clear, he nevertheless formulated his universal law of gravity without specifying how it was mediated.

Albert Einstein, building on the ideas of Maxwell, discovered a suitable mediating agent: a distortion of the structure of space-time itself. Einstein's contributions made classical physics into what is called a *local theory*: there is no action at a distance. All influences are transmitted essentially by contact interactions between tiny neighboring mathematically described "entities," and no influence propagates faster than the speed of light.

Classical physics is, moreover, *deterministic*: the interactions are such that the state of the physical world at any time is completely determined by the state at any earlier time. Consequently, according to classical theory, the complete history of the physical world *for all time* is mechanically fixed by contact interactions between tiny component parts, together with the initial condition of the primordial universe.

This result means that, according to classical physics, *you are a mechanical automaton*: your every physical action was pre-determined before you were born solely by mechanical interactions between tiny mindless entities. Your mental aspects are *causally redundant*: everything you do is completely determined by mechanical conditions alone, without reference to your thoughts, ideas, feelings, or intentions. Your intuitive feeling that your mental intentions make a difference in what you do is, according to the principles of classical physics, a false and misleading illusion.

There are two possible ways within classical physics to understand this total incapacity of your mental side (i.e., mental processes and consciousness) to make any difference in what you do. The first way is to consider your thoughts, ideas, and feelings to be epiphenomenal by-products of the activity of your brain. Your mental side is then a causally impotent sideshow that *is produced*, or *caused*, by your brain, but that *produces* no reciprocal action back upon your brain. The second way is to contend that each of your conscious experiences --- each of your thoughts, ideas, or feelings --- is the *very same thing* as some pattern of motion of various tiny parts of your brain.

## 4. Problems with classical physics

William James (1890: 138) argued against the first possibility, epiphenomenal consciousness, by claiming that "*The particulars of the distribution of consciousness*, so far as we know them, *points to its being efficacious*." He noted that consciousness seems to be "an organ, superadded to the other organs which maintain the animal in its struggle for existence; and the presumption of course is that it helps him in some way in this struggle, just as they do. But it cannot help him without being in some way efficacious and influencing the course of his bodily history." James said that the study described in his book "will show us that consciousness is at all times primarily a *selecting agency*." It is present when choices must be made between different possible courses of action. He further mentioned that "It is to my mind quite inconceivable that consciousness should have *nothing to do* with a business to which it so faithfully attends."(1890: 136)

If mental processes and consciousness have no effect upon the physical world, then what keeps a person's mental world aligned with his physical situation? What keeps his pleasures in general alignment with actions that benefit him, and pains in general correspondence with things that damage him, if pleasure and pain have no effect at all upon his actions?

These liabilities of the notion of epiphenomenal mind and consciousness lead most thinkers to turn to the alternative possibility that a person's mind and stream of consciousness is *the very same thing* as some activity in his brain: mind and consciousness are "emergent properties" of brains.

A huge philosophical literature has developed arguing for and against this idea. The primary argument against this "emergent-identity theory" position, *within a classical* *physics framework*, is that in classical physics the full description of nature is in terms of numbers assigned to tiny space-time regions, and there appears to be no way to understand or explain how to get from such a restricted conceptual structure, which involves such a small part of the world of experience, to the whole. How and why should that extremely limited conceptual structure, which arose basically from idealizing, by miniaturization, certain features of observed planetary motions, suffice to explain the totality of experience, with its pains, sorrows, hopes, colors, smells, and moral judgments? Why, *given the known failure of classical physics at the fundamental level*, should that richly endowed whole be explainable in terms of such a narrowly restricted part?

The core ideas of the arguments in favor of an identity-emergent theory of mind and consciousness are illustrated by Roger Sperry's example of a "wheel." (Sperry 1992) A wheel obviously does something: it is causally efficacious; it carries the cart. It is also an *emergent property*: there is no mention of "wheelness" in the formulation of the laws of physics, and "wheelness" did not exist in the early universe; "wheelness" *emerges* only under certain special conditions. And the macroscopic wheel exercises "top-down" control of its tiny parts. All these properties are perfectly in line with classical physics, and with the idea that "a wheel is, precisely, a structure constructed out of its tiny atomic parts." So why not suppose mind and consciousness to be, like "wheelness", emergent properties of their classically conceived tiny physical parts?

The reason that mind and consciousness are not analogous to "wheelness", within the context of classical physics, is that the properties that characterize "wheelness" are properties that are *entailed,* within the conceptual framework of classical physics, by properties specified in classical physics, whereas the properties that characterize conscious mental processes, namely the way it feels, are not *entailed,* within the conceptual structure provided by classical physics, by the properties specified by classical physics.

That is the huge difference-in-principle that distinguishes mind and consciousness from things that, according to classical physics, are constructible out of the particles that are postulated to exist by classical physics.

Given the state of motion of each of the tiny physical parts of a wheel, as it is conceived of in classical physics, the properties that characterize the wheel - e.g., its roundness, radius, center point, rate of rotation, etc., - are specified within the conceptual framework provided by the principles of classical physics, which specify only geometric-type properties such as changing locations and shapes of conglomerations of particles, and numbers assigned to points in space. But given the state of motion of each tiny part of the brain, as it is conceived of in classical physics, the properties that characterize the stream of consciousness - the painfulness of the pain, the feeling of the anguish, or of the sorrow, or of the joy - are not specified, within the conceptual framework provided by the principles of classical physics. Thus it is possible, within that classical physics framework, to strip away those feelings without disturbing the physical descriptions of the motions of the tiny parts. One can, within the conceptual framework of classical physics, take away the consciousness while leaving intact the properties that enter into that theoretical construct, namely the locations and motions of the tiny physical parts of the brain and its physical environment. But one cannot, within the conceptual framework provided by classical physics, take away the "wheelness" of a wheel without affecting the locations and motions of the tiny physical parts of the wheel.

Because one can, within the conceptual framework provided by classical physics, strip away mind and consciousness without affecting the physical behavior, one cannot rationally claim, *within that framework*, that mind and consciousness are the *causes* of the physical behavior, or are *causally efficacious* in the physical world. Thus the "identity theory" or "emergent property" strategy fails in its attempt to make mind and consciousness efficacious, within the conceptual framework provided by classical physics. Moreover, the whole endeavor to base brain theory on classical physics is undermined by the fact that the classical theory fails to work for phenomena that depend critically upon the properties of the atomic constituents of the behaving system, and brains are such systems: brain processes depend critically upon synaptic processes, which depend critically upon ionic processes that are highly dependent upon their quantum nature. This essential involvement of quantum effects will be discussed in detail in a later section.

## 5. The Quantum approach

Classical physics is *an approximation* to a more accurate theory - called quantum mechanics - and quantum mechanics makes mind and consciousness efficacious. Quantum mechanics *explains* the causal effects of mental intentions upon physical systems: it explains how your mental effort can influence the brain events that cause your body to move. Thus quantum theory converts science's picture of you from that of a mechanical automaton to that of a mindful human person. Quantum theory also shows, explicitly, how the approximation that reduces quantum theory to classical physics completely eliminates the quantum mechanically described effects of your conscious thoughts upon your brain and body. Hence, from a physics point of view, trying to understand the connection between mind/consciousness and brain by going to the classical approximation is absurd: it amounts to trying to understand something in an approximation that eliminates the effect you are trying to study.

Quantum mechanics arose during the twentieth century. Scientists discovered, empirically, that the principles of classical physics were not correct. Moreover, they were wrong in ways that no minor tinkering could ever fix. The *basic principles* of classical physics were thus replaced by *new basic principles* that account uniformly both for all the successes of the older classical theory and also for all the newer data that is incompatible with the classical principles.

The most profound alteration of the fundamental principles was to bring the mind and consciousness of human beings into the basic structure of the physical theory. In fact, the whole *conception of what science is* was turned inside out. The core idea of classical physics was to describe the "world out there," with no reference to "our thoughts in here." But the core idea of quantum mechanics is to describe *our activities as knowledge-seeking agents*, and *the knowledge that we thereby acquire*. Thus quantum theory involves, basically, what is "in here," not just what is "out there."

The basic philosophical shift in quantum theory is the *explicit* recognition that science is about *what we can know*. It is fine to have a beautiful and elegant mathematical theory about a *really existing physical world out there* that meets a lot of intellectually satisfying criteria. But the essential demand of *science* is that the theoretical constructs be tied to the experiences of the human scientists who devise ways of testing the theory, and of the human engineers and technicians who both participate in these tests, and eventually put the theory to work. So the structure of a proper physical theory must involve not only the part describing the behavior of the not-directly-experienced theoretically postulated entities, expressed in some appropriate symbolic language, but also a part describing the human experiences that are pertinent to these tests and applications, expressed in the language that we actually use to describe such experiences to ourselves and to each other. Finally we need some "bridge laws" that specify the connection between the concepts described in these two different languages.

Classical physics met these requirements in a rather trivial kind of way, with the relevant experiences of the human participants being taken to be direct apprehensions of gross behaviors of large-scale properties of big objects composed of huge numbers of the tiny atomic-scale parts. These apprehensions --- of, for example, the perceived location and motion of a falling apple, or the position of a pointer on a measuring device --- were taken to be *passive*: they had no effect on the behaviors of the systems being studied. But the physicists who were examining the behaviors of systems that depend sensitively upon the behaviors of their tiny atomic-scale components found themselves forced to go to a less trivial theoretical arrangement, in which the human agents were no longer passive observers, but were *active participants* in ways that contradicted, and were impossible to comprehend within, the general framework of classical physics, *even when the only features of the physically described world that the human beings observed were large-scale properties of measuring devices*. The sensitivity of the behavior of the devices to the behavior of some tiny atomic-scale particles propagates to devices and observers in such a way that the choice made by an observer about what sort of knowledge to seek can profoundly affect the knowledge that can ever be received either by that observer himself or by any other observer with whom he can communicate. Thus, for all practical purposed, the *choice* made by the observer about how he will act *affects* the physical system being acting upon. That itself is not the least bit surprising: how one acts on a tiny system would certainly be expected to affect it. Nor is it shocking that the exact form of this effect is specified in quantum mechanics by precise mathematical rules. But the key point should not be overlooked: the logical structure of the basic physical theory has become profoundly transformed. The connection between the mathematically specified physical properties of a system and *the agent's choice of which item of knowledge about that system is to be extracted* is changed from one in which the agent's choice has no effect at all on the mathematically described system to one in which that choice has a very specific mathematically described effect on that mathematically described system. This revision of the relationship between *knowledge-related choices* and *mathematical descriptions of physical properties* might be expected to have ramifications in neuroscience, in situations where the causal effects of knowledge-related *choices* is at issue.

This original formulation of quantum theory was created mainly at an Institute in Copenhagen directed by Niels Bohr, and is called "The Copenhagen Interpretation." Due to the puzzling strangeness of the properties of nature entailed by the new mathematics, the Copenhagen strategy was to refrain from making any ordinary sort of ontological claims, but to take, instead, an essentially pragmatic stance. Thus the theory was formulated *basically* as a set of practical rules for how scientists should go about the task of acquiring useful knowledge pertaining to the environment in which they were somehow imbedded, and then using this knowledge in practical ways. Claims about "what the world out there is really like" were considered to lie outside of *science* if they make no practical difference.

The most profound change in the principles is encapsulated in Niels Bohr dictum that "in the great drama of existence we ourselves are both actors and spectators." (Bohr 1963: 15 and 1958: 81) The emphasis here is on "actors": in classical physics we were mere spectators.

Copenhagen quantum theory is about the relationships between human agents (called *participants* by John Wheeler) and the systems upon which they act. In order to achieve this conceptualization the Copenhagen formulation separates the physical universe into two parts, which are described in two different languages. One part is the observing human agent and his measuring devices. This extended "agent," which includes the devices, is described in mental terms - in terms of our instructions to colleagues about how to set up the devices, and our reports of what we then "see," or otherwise consciously experience. The other part of nature is *the system that the "agent" is acting upon*. That part is described in physical terms - in terms of mathematical properties assigned to tiny space-time regions. Thus Copenhagen quantum theory brings "doing science" into science. In particular, it brings a crucial part of doing science, namely our *choices* about how to probe physical systems, directly into the causal structure. And it specifies the non-trivial, and not classically understandable, effects of these choices upon the systems being probed.

This approach works very well in practice. However, it seems apparent that the body and brain of the human agent, and his devices, are parts of the physical universe, and hence that a complete theory ought to be able to describe also our bodies and brains in physical terms. On the other hand, the structure of the theory centrally involves also the empirical realities described in mentalistic language as our intentional probing actions and the resulting experiential feedbacks.

The great mathematician and logician John von Neumann carefully formulated the theory in a rigorous way that allows the bodies and brains of the agents, along with their measuring devices, to be placed in the physically described world, *while retaining those mentalistically described actions made by the agents that are central to the theory.*

Von Neumann identifies two very different processes that enter into the quantum theoretical description of the evolution of a physical system. He calls them Process 1 and Process 2 (von Neumann 1955: 418). Process 2 is the analog in quantum theory of the process in classical physics that takes the state of a system at one time to its state at a later time. This Process 2, like its classical analog, is *local* and *deterministic*. However, Process 2 by itself is not the whole story: it generates "physical worlds" that do not agree with human experiences. For example, if Process 2 were, from the time of the Big Bang, the *only* process in nature, then the quantum state of the moon would represent a structure smeared out over large part of the sky, and each human body-brain would likewise be represented by a structure smeared out continuously over a huge region.

To tie the quantum mathematics to human experience in a rationally coherent and mathematically specified way quantum theory invokes *another process*, which von Neumann calls Process 1.

Any physical theory must, in order to be complete, specify how the elements of the theory are connected to human experience. In classical physics this connection is part of a *metaphysical* superstructure: it is not part of the core dynamical description. But in quantum theory this connection of the mathematically described physical state to human experiences is placed *within the causal structure.* And this connecting process is not passive: it does not represent a mere *witnessing* of a physical feature of nature by a passive mind. Rather, the process is active: *it injects into the physical state of the system being acted upon specific properties that depend upon the choices made by the agent.*

Quantum theory is built upon the practical concept of intentional actions by agents. Each such action is expected or intended to produce an experiential response or feedback. For example, a scientist might act to place a Geiger counter near a radioactive source, and expect to see the counter either "fire" during a certain time interval or not "fire" during that interval. The experienced response, "Yes" or "No", to the question "Does the counter fire during the specified interval?" specifies one bit of information. Quantum theory is thus an information-based theory built upon the information-acquiring actions of agents, and the information that these agents thereby acquire.

Probing actions of this kind are performed not only by scientists. Every healthy and alert infant is engaged in making willful efforts that produce experiential feedbacks, and he/she soon begins to form expectations about what sorts of feedbacks are likely to follow from some particular kind of effort. Thus both empirical science and normal human life are based on paired realities of this action-response kind, and our physical and psychological theories are both basically attempting to understand these linked realities within a rational conceptual framework.

The basic building blocks of quantum theory are, then, a set of intentional actions by agents, and for each such action an associated collection of possible "Yes" feedbacks, which are the possible responses that the agent can judge to be in conformity to the criteria associated with that intentional act. For example, the agent is assumed to be able to make the judgment "Yes" the Geiger counter clicked or "No" the Geiger counter did not click. Science would be difficult to pursue if scientists could make no such judgments about what they were experiencing.

All known physical theories involve idealizations of one kind or another. In quantum theory the main idealization is not that every object is made up of miniature planet-like objects. It is rather that there are agents that perform intentional acts each of which can result in a feedback that may or may not conform to a certain criterion associated with that act. One bit of information is introduced into the world in which that agent lives, according to whether the feedback conforms or does not conform to that criterion. Thus, knowing whether the counter clicked or not places the agent on one or the other of two alternative possible separate branches of the course of world history.

These remarks reveal the enormous difference between classical physics and quantum physics. In classical physics the elemental ingredients are tiny invisible bits of matter that are idealized miniaturized versions of the planets that we see in the heavens, and that move in ways unaffected by our scrutiny, whereas in quantum physics the elemental ingredients are intentional actions by agents, the feedbacks arising from these actions, and the effects of these actions upon the physically described states of the probed systems.

An intentional action by a human agent is partly an intention, described in psychological terms, and partly a physical action, described in physical terms. The feedback also is partly psychological and partly physical. In quantum theory these diverse aspects are all represented by logically connected elements in the mathematical structure that emerged from the seminal discovery of Heisenberg. That discovery was that in order to get a satisfactory quantum generalization of a classical theory one must replace various *numbers* in the classical theory by *actions* (operators). A key difference between numbers and actions is that if A and B are two actions then AB represents the action obtained by performing the action A upon the action B. If A and B are two different actions then generally AB is different from BA: the order in which actions are performed matters. But for numbers the order does not matter: AB = BA.

The difference between quantum physics and its classical approximation resides in the fact that in the quantum case certain differences AB-BA are proportional to a number measured by Max Planck in 1900, and called Planck's constant. Setting those differences to zero gives the classical approximation. Thus quantum theory is closely connected to classical physics, but is incompatible with it, because certain *nonzero* quantities must be replaced by zero to obtain the classical approximation.

The intentional actions of agents are represented mathematically in Heisenberg's space of actions. Here is how it works.

Each intentional action depends, of course, on the *intention of the agent*, and upon the *state of the system* upon which this action acts. Each of these *two aspects of nature* is represented within Heisenberg's space of actions by an *action*. The idea that a "state" should be represented by an "action" may sound odd, but Heisenberg's key idea was to replace what classical physics took to be a "being" by a "doing." I shall denote the action that represents the state being acted upon by the symbol S.

An intentional act is an action that is intended to produce a feedback of a certain conceived or imagined kind. Of course, no intentional act is sure-fire: one's intentions may not be fulfilled. Hence the intentional action puts in play a process that will lead either to a confirmatory feedback "Yes," the intention is realized, or to the result "No", the "Yes" response did not occur.

The effect of this intentional mental act is represented mathematically by an equation that is one of the key components of quantum theory. This equation represents, within the quantum mathematics, the effect of the Process 1 action upon the quantum state S of the system being acted upon. The equation is:

$$S \rightarrow S' = PSP + (I-P) S(I-P).$$

This formula exhibits the important fact that this Process 1 action changes the state S of the system being acted upon into a new state S', which is a sum of two parts.

The first part, PSP, represents the possibility in which the experiential feedback called "Yes" appears, and the second part, (I-P) S(I-P), represents the alternative possibility "No", this "Yes" feedback does not appear. Thus an effect of the probing action is injected into the mathematical description of the physical system being acted upon.

The operator P is important. The action represented by P, acting both on the right and on the left of S, is the action of eliminating from the state S all parts of S except the "Yes" part. That particular retained part is determined by the choice made by the agent. The symbol I is the unit operator, which is essentially multiplication by the number 1, and the action of (I-P), acting both on the right and on the left of S, is, analogously, to eliminate from S all parts of S except the "No" parts.

Notice that Process 1 produces the *sum* of the two alternative possible feedbacks, not just one or the other. Since the feedback must either be "Yes" or "No = Not-Yes," one might think that Process 1, which keeps both the "Yes" and the "No" possibilities, would do nothing. But that is not correct! This is a key point. It can be made quite clear by noticing that S can be written as a sum of four parts, only two of which survive the Process 1 action:

$$S = PSP + (I-P) S(I-P) + PS(I-P) + (I-P)SP.$$

This formula is a strict identity. The dedicated reader can quickly verify it by collecting the contributions of the four occurring terms PSP, PS, SP, and S, and verifying that all terms but S cancel out. This identity shows that the state S is a sum of four parts, *two of which are eliminated by Process 1*.

But this means that Process 1 has a *nontrivial effect* upon the state being acted upon: it eliminates the two terms that correspond neither to the appearance of a "Yes" feedback nor to the failure of the "Yes" feedback to appear.

That is the *first key point*: quantum theory has a specific causal process, Process 1, which produces a nontrivial effect of an *agent's choice* upon the physical description of the system being examined. ["Nature" will eventually choose between "Yes" and "No", but I focus here on the prior Process 1, *the agent's choice*. I call Nature's subsequent choice Process 3.]

## 5.1. Free choices

The second key point is this: the agent's choices are "free choices," *in the specific sense specified below*.

Orthodox quantum theory is formulated in a realistic and practical way. It is structured around the activities of human agents, who are considered able to freely elect to probe nature in any one of many possible ways. Bohr emphasized the freedom of the experimenters in passages such as:

> "The freedom of experimentation, presupposed in classical physics, is of course retained and corresponds to the free choice of experimental arrangement for which the mathematical structure of the quantum mechanical formalism offers the appropriate latitude." (Bohr 1958: 73}

This freedom of choice stems from the fact that in the original Copenhagen formulation of quantum theory the human experimenter is considered to stand outside the system to which the quantum laws are applied. Those quantum laws are the only precise laws of nature recognized by that theory. Thus, according to the Copenhagen philosophy, *there are no presently known laws that govern the choices* made by the agent/experimenter/observer about how the observed system is to be probed. This choice is*, in this very specific sense*, a "free choice."

An awkward feature of the Copenhagen formulation is that the devices, bodies, and brains, though consisting of, or containing, atomic particles, are excluded from the physically described world. This produces mathematical inconsistencies. Von Neumann resolves this problem by shifting the boundary between "observer" and "observee", in a series of steps, until finally all parts of the physical world, including all physically described bodies and brains, are in the physically described system. At each stage the directly observed part of the observed system is the part closest to the boundary. At the final stage this part is the brain of the agent, or some part of that brain.

For the "good measurement" cases amenable to study von Neumann showed that shifting the boundary preserved the validity of the quantum rules, thus ensuring the equivalence of von Neumann's rules in the limiting case to the pragmatically verified rules in the Copenhagen case. At no stage does the agent's choice become determined by Process 2: it remains "free", in the sense specified above. At the final stage, in which all bodies and brains become included in the physically described world, the tasks of the observer/agent devolve upon what von Neumann calls an "abstract ego." (von Neumann 1955: 421). Pragmatically, this is just the agent's stream of conscious events and conscious choices: it's the agent's mind.

A complete mind-brain theory must specify how brain affects mind. One of the authors (HPS) is endeavoring with K. Laskey to apply Bayesian decision theory to this problem (see Laskey 2004). But the present article is about the effect of mind on brain: it is about the causal effects of an agent's Process 1 conscious choices upon his or her brain.

## 5.2. Probabilities

The predictions of quantum theory are generally statistical: they specify, for each of the alternative possible feedbacks, only the *probability* that the agent will experience that feedback. Which of these alternative possible feedbacks will actually occur in response to the chosen Process 1 probing action is not determined by quantum theory.

The formula for the probability that the agent will experience the feedback 'Yes' is Tr PSP/Tr S, where the symbol Tr represents the trace operation. This trace operation means that the actions act in a cyclic fashion, so that the rightmost action acts back around upon the leftmost action. Thus, for example, Tr ABC = Tr CAB = Tr BCA. The product ABC represents the result of letting A act upon B, and then letting that product AB act upon C. But what does C act upon? Taking the trace of ABC means specifying that C acts back around on A.

An important property of a trace is that the trace of any of the sequences of actions that we consider must always give a positive number or zero. This trace operation is what ties the actions, as represented in the mathematics, to measurable numbers.

Von Neumann generated his form of quantum theory by first recognizing that, at the Copenhagen level, the Process 1 action S→S'= PSP + (I-P) S(I-P) describes an influence of a mentalistically describable choice upon a physically described state, and by then expanding, by a series of steps, the physically described portion of nature to include the state of the brain connected to that mentalistically described choice. Thus the Process 1 action comes to represents a dynamical influence of a free choice made by an agent upon his own brain. This means that orthodox (von Neumann) quantum theory has a Process 1 action that: (1) is needed to tie the theory to the experience of an agent, (2) is "freely chosen" by that agent, and (3) produces a specified effect on the state of the brain of that agent.

As previously mentioned, von Neumann called the mentalistically described aspect of the agent "his abstract 'ego'." This phrasing tends to conjure up the idea of a disembodied entity, standing somehow apart from the body/brain. But another possibility is that consciousness is an *emergent property* of the brain. Some of the problems that occur when trying to defend this idea of emergence within the framework of classical physical theory disappear in quantum theory. For one thing, there is now no need to defend against the charge that the emergent properties, mind and consciousness, play no needed role, because whatever they do physically is already done by matter alone. In quantum theory the Process 1 choices play a key role filled by nothing else. For another thing, the concept of matter was designed to be mindless: it is logically impossible to take the austere naked concept of physical particles that classical physicists build their imagined conception of the material world upon, and build out of them a conscious thought. On the other hand, the mathematically described quantum state of the universe is compendium of past knowledge that is the grist for present and future knowledge. There are no ontological qualities to draw upon besides knowledge, feelings and their forebearers. But the important thing is that those ontological questions are, just as in atomic physics, irrelevant to the practical applications of the theory. What matters is the structural relationships that the theory provides between the two kinds of data. As Heisenberg remarked:

"The conception of the objective reality of the elementary particles has thus evaporated not into the cloud of some obscure new reality concept, but into the transparent clarity of mathematics that represents no longer the behavior of the particle but rather our knowledge of this behavior." (Heisenberg 1958)

The quantum state of a human brain is a complex thing. But its main features can be understood by considering first a classical conception of the brain, and then incorporating some key features that arise already in the case of the quantum state associated with single degree of freedom, which could be the quantum analog of the center point of some large or small object, such as a planet or a calcium ion.

## 5.3. States of a simple harmonic oscillator

The most important example of a quantum state is that of a pendulum, or more precisely, what is called a "simple harmonic oscillator." Such a system is one in which there is a restoring force that tends to push the center of the object to a single "base point" of lowest energy, and in which the strength of this restoring force is directly proportional to the distance of the center point of the object from this base point.

According to classical physics any such system has a state of lowest energy. In this state the center point of the object lies motionless at the base point. In quantum theory this system again has a state of lowest energy, but the center point is not localized at the base point: the location of the center point is represented by a *cloudlike* spatial structure that is spread out over a region that extends to infinity. However, the amplitude of this cloudlike form has the shape of a bell: it is largest at the base point, and falls off in a prescribed manner as the distance of the center point from the base point increases.

If one were to squeeze this state of lowest energy into a more narrow space, and then let it loose, the cloudlike form would first explode outward, but then settle into an oscillating motion. Thus the cloudlike spatial structure behaves rather like a swarm of bees, such that the more they are squeezed in space the faster they move, and the faster the squeezed cloud will explode outward when the squeezing constraint is released. These visualizable properties extend in a natural way to many-particle cases. However, it should be emphasized that the "swarm of bees" analogy cannot be pushed too far, because the cloud like structure refers, in the simple one-particle case, to *one single particle* ---e.g., to one calcium ion --- isolated from all others. The different parts of the cloud that represents this *one single calcium ion*, seem to be repelling each other, in the case of the squeezed state.

5.4. The double-slit experiment

An important difference between the behavior of the quantum cloudlike form and the somewhat analogous *classical probability distribution* is exhibited by the famous *double-slit experiment*. If one shoots an electron, an ion, or any other quantum counterpart of a tiny classical object, at a narrow slit then if the object passes through the slit the associated cloudlike form will fan out over a wide angle. This is analogous to the initial explosion of the tightly confined swarm of bees. But if one opens two closely neighboring narrow slits, then what passes through the slits is described by a probability distribution that is not just the sum of the two separate fanlike structures that would be present if each slit were opened separately. Instead, at some points the probability value will be almost *twice the sum* of the values associated with the two individual slits, and in other places the probability value drops nearly to zero, even though both individual fanlike structures give a large probability value at that place. These *interference* features of the quantum cloudlike structure make that structure logically different from a classical-physics probability distribution---for a single particle --- because in the classical case the probabilities arising from the two slits would simply add, due to the facts that, according to classical principles, the single particle must pass through one slit or the other, and that the presence of the other opening would not matter much.

Quantum theory deals consistently with this interference effect, and all the other non-classical properties of these cloudlike structures.

5.5. Nerve terminals, ion channels, and the need to use quantum theory

Some neuroscientists who study the relationship of mind and consciousness to brain processes believe that classical physics will be adequate for that task. That belief would have been reasonable during the nineteenth century, but now, in the twenty-first century, it is rationally untenable: quantum theory must in principle be used because the behavior of the brain depends sensitively upon ionic and atomic processes, and these processes often involve large quantum effects.

To study quantum effects in brains within an orthodox (i.e., Copenhagen or von Neumann) quantum theory one must use the von Neumann formulation. The reason is that *Copenhagen* quantum theory is formulated in a way that leaves out the quantum dynamics of the human observer's body and brain. But von Neumann quantum theory takes the physical system S upon which the crucial Process 1 acts to be the brain of the agent. Thus Process 1 describes an interaction between a person's stream of consciousness, described in mentalistic terms, and the activity in his brain, described in physical terms. That interaction drops completely out when one passes to the classical approximation. Hence ignoring quantum effects in the study of the connection between mind/consciousness and brain means, according to the basic principles of physics, ignoring the dynamical connection one is trying to study. One must *in principle* use quantum theory. But there is then the quantitative issue of how important the quantum effects are.

To explore that question we consider the quantum dynamics of nerve terminals.

Nerve terminals are essential connecting links between nerve cells. The general way they work is reasonably well understood. When an action potential traveling along a nerve fiber reaches a nerve terminal, a host of ion channels open. Calcium ions enter through these channels into the interior of the terminal. These ions migrate from the channel exits to release sites on vesicles containing neurotransmitter molecules. A triggering effect of the calcium ions causes these contents to be dumped into the synaptic cleft that separates this terminal from a neighboring neuron, and these neurotransmitter molecules influence the tendencies of that neighboring neuron to "fire."

The channels through which the calcium ions enter the nerve terminal are called "ion channels." At their narrowest points they are less than a nanometer in diameter (Cataldi et al. 2002). This extreme smallness of the opening in the ion channels has profound quantum mechanical implications. The consequence is essentially the same as the consequence of the squeezing of the state of the simple harmonic operator, or of the narrowness of the slits in the double-slit experiments. The narrowness of the channel restricts the lateral *spatial* dimension. Consequently, the lateral *velocity* is forced by the *quantum uncertainty principle* to become large. This causes the cloud associated with the calcium ion to *fan out* over an increasing area as it moves away from the tiny channel to the target region where the ion will be absorbed as a whole, or not absorbed, on some small triggering site.

This spreading of the ion wave packet means that the ion may or may not be absorbed on the small triggering site. Accordingly, the vesicle may or may not release its contents. Consequently, the quantum state of the vesicle has a part in which the neurotransmitter is released and a part in which the neurotransmitter is not released. This quantum splitting occurs at every one of the trillions of nerve terminals.

It would seem at first that this splitting of the state of the brain into alternative possibilities at each possible release of each possible vesicle would create an incomprehensible situation, and that one must appeal to the fact that in a warm wet brain various decoherence effects will enter, and tend to wash out all quantum effects beyond local chemical processes, which can be conceived to be imbedded in an essentially classical world. Strong decoherence effects of this kind are certainly present, and are described by the von Neumann formulation employed here. These effects convert the state S of the brain into what is called a "statistical mixture" of "nearly classically describable" states, each of which develops in time, in the absence of Process 1 events, in an almost classically describable way.

*This decoherence effect makes the main consequences of quantum theory easily accessible to neuroscientists by effectively reducing the complex quantum state of the brain to collection of almost classically describable possibilities.* Because of the uncertainties introduced at the ionic level, the brain state will develop not into one single classically describable macroscopic state, but into a continuous distribution of parallel virtual states of this kind. It is Process 1 that, as will now be described, allows definite empirical predictions to be extracted from this continuous collection of parallel overlapping almost-classical possibilities.

A principal function of the brain is to receive clues from the environment, to form an appropriate plan of action, and to direct and monitor the activities of the brain and body specified by the selected plan of action. The exact details of the plan will, for a classical model, obviously depend upon the exact values of many noisy and uncontrolled variables. In cases close to a bifurcation point the dynamical effects of noise might even tip the balance between two very different responses to the given clues, e.g., tip the balance between the 'fight' or 'flight' response to some shadowy form.

The effect of the independent "release" or "don't release" options at each of the trigger sites, coupled with the uncertainty in the timing of the vesicle release at each of the trillions of nerve terminals will be to cause the quantum mechanical state of the brain to become a smeared out cloud of different macroscopic possibilities, some representing different alternative possible plans of action. As long as the brain dynamics is controlled wholly by Process 2 - which is the quantum generalization of the Newtonian laws of motion of classical physics - all of the various alternative possible plans of action will exist in parallel, with no one plan of action singled out as the one that will actually be experienced.

Some process beyond the local deterministic Process 2 is required to pick out one experienced course of physical events from the smeared out mass of possibilities generated by all of the alternative possible combinations of vesicle releases at all of the trillions of nerve terminals. This other process is Process 1. It brings in a *choice* that is not determined by any currently known law of nature, yet has a definite effect upon the brain of the chooser. The choice must pick an operator P, and also a time t at which P acts. The effect of this action is to change the state $S(t)$ of the brain, or of some large part of the brain, to $PS(t)P + (I-P) S(t) (I-P)$.

The action P cannot act at a *point* in the brain, because a point action would dump a huge (in principle infinite) amount of energy into the brain, which would then explode. The operator P must therefore act non-locally, over a potentially large part of the brain.

To obtain a satisfactory theory, the operators P must involve a completely different set of variables. The pertinent variables for Process 1 are not the coordinates of the various individual calcium ions, but rather certain *quasi-stable macroscopic degrees of freedom*. The selected brain structures must enjoy the stability, endurance, and causal connections needed to bring into being the intended experiential feedbacks.

These structures are likely to be more like the lowest-energy state of the simple harmonic oscillator discussed above, which is stable, or like the states obtained from such lowest-energy states by spatial displacements and shifts in velocity. These states tend to endure as oscillating states, rather than immediately exploding. In other words, in order to get the needed causal structure the projection operators P corresponding to intentional actions ought to be constructed out of *oscillating states of macroscopic subsystems of the brain*, rather than out of the states of the individual particles. The states associated with Process 1 would then be functionally important brain analogs of collections of oscillating modes of a drumhead, in which large assemblies of particles of the brain are moving in a coordinated way that will lead on, via the mechanical laws, to further coordinated activities.

The brain process that is actualized by the transition $S(t) \rightarrow PS(t)P$ is the neural correlate of the psychological intended action. It is the brain's template for the intended action.

## 5.6. Choices of the Process 1 actions

It has been emphasized that the choices of which Process I actions actually occur are "free choices," in the sense that they are not specified by the currently known laws of physics. On the other hand, a person's intentions are surely related in some way to his historical past. This means that the laws of contemporary orthodox quantum theory, although restrictive and important, are not the whole story. However, orthodox quantum theory, although making no claim to ontological completeness, achieves a certain pragmatic completeness by ascribing the Process 1 choices to the will of the psychologically described agent. An ontologically complete theory must do better, but, in keeping with quantum theory, we endeavor to achieve practical utility by exploiting the established laws themselves, without making a detailed commitment or theory about the origin of will. *Willful choices become empirical inputs rather than mechanical effects.*

It is useful to classify Process I events as either "active" or "passive." The *passive* Process I events are considered to occur automatically, in accordance with some brain-controlled rule, with little or no involvement of conscious effort. The *active* Process I events are intentional and involve effort.

Consciousness probably contributes very little to brain dynamics, compared to the contribution of the brain itself. To minimize the input of consciousness, and in order to achieve testability, we propose to allow mental effort to do nothing but increase the "density of attention", which is a measure of the *rapidity* of the events in the Process 1 sequence. This allows mental effort to have only a very limited kind of influence on brain activities that are largely controlled by the brain itself.

The notion that only the attention density was controlled by conscious effort arose from an investigation into what sort of conscious control over Process 1 action was necessary and sufficient to accommodate the most blatant empirical facts. Imposing this strong limitation on the allowed effects of consciousness produces a theory with correspondingly strong predictive power. In this model all significant effects of consciousness upon the brain arise exclusively from a well known and well verified strictly quantum effect known as the Quantum Zeno Effect.

## 5.7. The Quantum Zeno effect

If one considers only passive events, then it is very difficult to identify any empirical effect of Process 1, apart from the occurrence of awareness. In the first place, the empirical averaging over the "Yes" and "No" possibilities tends to wash out all effects that depart from what would arise from a classical statistical analysis that incorporates the uncertainty principle as simply lack of knowledge. Moreover, the passivity of the mental process means that we have no empirically controllable variable.

But the study of effortfully controlled intentional action brings in two empirically accessible variables, the intention and the amount of effort. It also brings in the important physical Quantum Zeno Effect. This effect is named for the Greek philosopher Zeno of Elea, and was brought into prominence in 1977 by the physicists Sudarshan and Misra (1977). It gives a name to the fact that repeated and closely-spaced observational acts can effectively hold the "Yes" feedback in place for an extended time interval that depends upon the *rapidity at which the Process I actions are happening*. According to our model, this rapidity is controlled by the amount of effort being applied. In our notation the effect is to keep the "Yes" state PS(t)P associated with the intended experiential response in place longer than would be the case if no effort were being made. This effect can override very strong mechanical forces arising from Process 2. It's a case of mind over (brain) matter! The "Yes" state PS(t)P is conditioned by training and learning to contain the template for action which if held in place for an extended period will tend to produce the intended feedback. Thus the model allows mental effort to tend to bring intended experiences into being. Moreover, systems that have the capacity to exploit this convenient feature of quantum theory obviously enjoy a tremendous advantage over systems that do not or cannot exploit it.

## 6. Support from psychology

A person's experiential life is a stream of conscious experiences. The person's experienced *"self"* is *part* of this stream of consciousness: it is not an extra thing that is outside or apart from the stream. In James's words "*thought is itself the thinker*, and psychology need not look beyond." The "self" is a slowly changing "fringe" part of the stream of consciousness. It provides a background cause for the central focus of attention.

The physical brain, evolving mechanically in accordance with the local deterministic Process 2 can do most of the necessary work of the brain. It can do the job of creating, on the basis of its interpretation of the clues provided by the senses, a suitable response, which will be controlled by a certain pattern of neural or brain activity that acts as a *template for action*. But, due to its quantum nature, the brain necessarily generates an amorphous mass of overlapping and conflicting templates for action. Process 1 acts to extract from this jumbled mass of possibilities some particular template for action. This is the preferred "Yes" state PSP that specifies the form of the Process 1 event. But the quantum rules do not assert that this "Yes" part of the prior state S *necessarily* comes into being. They assert, instead, that if this Process 1 action is triggered---say by some sort of "consent"---then this "Yes" component PSP will come into being with probability Tr PSP/Tr S, and that the "No" state will occur if the "Yes" state does not occur.

If the rate at which these "consents" occur is assumed to be increasable by conscious mental effort, then the causal efficacy of "will" can be understood. Conscious effort can, by activation of the Quantum Zeno Effect, override strong mechanical forces arising from Process 2, and cause the template for action to be held in place longer than it would if the rapid sequence of Process 1 events were not occurring. This sustained existence of the template for action can cause that action to occur.

Does this quantum-physics-based conception of the origin of the causal efficacy of "Will" accord with the findings of psychology?

Consider some passages from "Psychology: The Briefer Course", written by William James. In the final section of the chapter on attention James(1892: 227) writes:

> "I have spoken as if our attention were wholly determined by neural conditions. I believe that the array of things we can attend to is so determined. No object can catch our attention except by the neural machinery. But the amount of the attention which an object receives after it has caught our attention is another question. It often takes effort to keep mind upon it. We feel that we can make more or less of the effort as we choose. If this feeling be not deceptive, if our effort be a spiritual force, and an indeterminate one, then of course it contributes coequally with the cerebral conditions to the result. Though it introduces no new idea, it will deepen and prolong the stay in consciousness of innumerable ideas which else would fade more quickly away."

In the chapter on will, in the section entitled "Volitional effort is effort of attention" James (1892: 417) writes:

> "Thus we find that we reach the heart of our inquiry into volition when we ask by what process is it that the thought of any given action comes to prevail stably in the mind."

and later

"The essential achievement of the will, in short, when it is most 'voluntary,' is to attend to a difficult object and hold it fast before the mind. ... Effort of attention is thus the essential phenomenon of will."

Still later, James says:

"Consent to the idea's undivided presence, this is effort's sole achievement."... ``Everywhere, then, the function of effort is the same: to keep affirming and adopting the thought which, if left to itself, would slip away."

This description of the effect of will on the course of mental-cerebral processes is remarkably in line with *what had been proposed independently from purely theoretical considerations of the quantum physics of this process*. The connections specified by James are *explained* on the basis of the same dynamical principles that had been introduced by physicists to explain atomic phenomena. Thus the whole range of science, from atomic physics to mind-brain dynamics, is brought together in a single rationally coherent theory of an evolving cosmos that is constituted not of matter but of actions that determine propensities or tendencies for Process 1 events to occur, and within which conscious agents could naturally evolve in accordance with the principles of natural selection, due to the fact that their conscious efforts have physical consequences.

In the quantum theory of mind/consciousness-brain being advocated here there are altogether three processes. First, there is the purely mechanical process called Process 2.

As discussed at length in the book, *Mind, Matter, and Quantum Mechanics* (Stapp 1993/2003: 150), this process, as it applies to the brain, involves important dynamical units that are represented by complex patterns of neural activity (or, more generally, of brain activity) that are "facilitated" (i.e., strengthened) by use, and are such that each unit tends to be activated as a whole by the activation of several of its parts. The activation of various of these complex patterns by cross referencing---i.e., by activation of several of its parts---coupled to feed-back loops that strengthen or weaken the activities of appropriate processing centers, appears to account for the essential features of the mechanical part of the dynamics in a way that often is not greatly different from that of a classical model, except for the entry of a host of parallel possibilities that according to the classical concepts cannot exist simultaneously.

The second process, von Neumann's Process 1**,** is needed in order to pick out from a chaotic continuum of overlapping possibilities some particular discrete possibility and its complement (The complement can be further divided, but the essential action is present in the choice of a particular "Yes" state PSP from the morass of possibilities in which it is imbedded). The third process is Nature's choice between "Yes" and "No." Nature's choice conforms to a statistical (propensity) rule, but the agent's choice has no constraint of any kind in the theory put forth by von Neumann.

Process 1 has itself two modes. The first is passive and can produce temporally isolated events. The second involves mental effort, and a rapid sequence of Process 1 events that bring importantly into play the Quantum Zeno Effect. The passive process can exploit the massively parallel processing capacities of Process 2, whereas the second mode involves an effortfully sustained rapid linear sequence of Process 1 events.

Active Process 1 intervention has, according to the quantum model described here, a distinctive form. It consists of a sequence of intentional actions, the rapidity of which can be increased with effort. Such an increase in Attention Density, defined as an increase in the number of observations per unit time, can bring into play the Quantum Zeno Effect, which tends to hold in place both those aspects of the state of the brain that are fixed by the sequence of intentional actions, and also the felt intentional focus of these actions. Attention Density is not controlled by any physical rule of orthodox contemporary quantum theory, but is taken both in orthodox theory and in our model to be subject to subjective volitional control. This concordance of atomic physics and neurodynamics is the core of our model.

## 6.1. Support from psychology of attention

A huge amount of empirical work on attention has been done since the nineteenth century writings of William James. Much of it is summarized and analyzed in Harold Pashler's 1998 book "*The Psychology of Attention*." Pashler organizes his discussion by separating perceptual processing from post-perceptual processing. The former type covers processing that, first of all, identifies such basic physical properties of stimuli as location, color, loudness, and pitch, and, secondly, identifies stimuli in terms of categories of meaning. The post-perceptual process covers the tasks of producing motor actions and cognitive action beyond mere categorical identification. Pashler emphasizes [p. 33] that the empirical "findings of attention studies… argue for a distinction between perceptual attentional limitations and more central limitations involved in thought and the planning of action." The existence of these two different processes with different characteristics is a principal theme of Pashler's book [e.g., pp. 33, 263, 293, 317, 404].

A striking difference that emerges from the analysis of the many sophisticated experiments is that the perceptual processes proceed essentially in parallel, whereas the post-perceptual processes of planning and executing actions form a single queue. This is in line with the distinction between "passive" and "active" processes. The former are essentially a passive stream of essentially one-shot Process 1 events, whereas the "active" processes involve effort-induced rapid sequences of Process 1 events that can saturate a given capacity. This idea of a limited capacity for serial processing of effort-based inputs is the main conclusion of Pashler's book. It is in accord with the quantum-based model, supplemented by the condition that there is a limit to how many effortful Process 1 events per second a person can produce.

Examination of Pashler's book shows that this quantum model accommodates naturally all of the complex structural features of the empirical data that he describes. Of key importance is his Chapter Six, in which he emphasizes a specific finding: strong empirical evidence for what he calls a central processing bottleneck associated with the attentive selection of a motor action. This kind of bottleneck is what the quantum-physics-based theory predicts: the bottleneck is precisely the single linear sequence of mind-brain quantum events that von Neumann quantum theory describes.

Pashler [p. 279] describes four empirical signatures for this kind of bottleneck, and describes the experimental confirmation of each of them. Much of part II of Pashler's book is a massing of evidence that supports the existence of a central process of this general kind.

The queuing effect is illustrated in a nineteenth century result described by Pashler: mental exertion reduces the amount of physical force that a person can apply. He notes that "This puzzling phenomenon remains unexplained." [p. 387]. However, it is an automatic consequence of the physics-based theory: creating physical force by muscle contraction requires an effort that opposes the physical tendencies generated by the Schröedinger equation (Process 2). This opposing tendency is produced by the Quantum Zeno Effect (QZE), and is roughly proportional to the number of bits per second of central processing capacity that is devoted to the task. So if part of this processing capacity is directed to another task, then the applied force will diminish.

The important point here is that there is in principle, in the quantum model, an essential dynamical difference between the unconscious processing carried out by the Schröedinger evolution, which generates via a local process an expanding collection of classically conceivable experiential possibilities, and the process associated with the sequence of conscious events that constitute the willful selection of an action. The former are not limited by the queuing effect, because Process 2 simply develops all of the possibilities in parallel: it is the Process 1 events that, in the von Neumann formulation, form a single temporal sequence.

The experiments cited by Pashler all seem to support this clear prediction of the quantum approach. It is important to note that this bottleneck is not automatic within classical physics. A classical model could easily produce, simultaneously, two responses in different modalities, say vocal and manual, to two different stimuli arriving via two different modalities, say auditory and tactile: the two processes could proceed via dynamically independent routes. Pashler [p. 308] notes that the bottleneck is undiminished in split-brain patients performing two tasks that, at the level of input and output, seem to be confined to different hemispheres. This could be accounted for by the non-local character of the projection operator P.

An interesting experiment mentioned by Pashler involves the simultaneous tasks of doing an IQ test and giving a foot response to a rapidly presented sequence of tones of either 2000 or 250 Hz. The subject's mental age, as measured by the IQ test, was reduced from adult to 8 years [p. 299]. This result supports the prediction of quantum theory that the bottleneck pertains to both "intelligent" behavior, which requires complex effortful processing, and the simple willful selection of a motor response.

Pashler also notes [p. 348] that "Recent results strengthen the case for central interference even further, concluding that memory retrieval is subject to the same discrete processing bottleneck that prevents simultaneous response selection in two speeded choice tasks."

In the section on "Mental Effort" Pashler reports [p.383] that "incentives to perform especially well lead subjects to improve both speed and accuracy", and that the motivation had "greater effects on the more cognitively complex activity". This is what would be expected if incentives lead to effort that produces increased rapidity of the events, each of which injects into the physical process, via quantum selection and reduction, bits of control information that reflect mental evaluation. Pashler notes [p.385] "Increasing the rate at which events occur in experimenter-paced tasks often increases effort ratings without affecting performance. Increasing incentives often raises workload ratings and performance at the same time." All of these empirical connections are in line with the general principle that effort increases Attention Density, with an attendant increase in the rate of directed conscious events, each of which inputs a mental evaluation and a selection or focusing of a course of action.

Additional supporting evidence comes from the studies of the stabilization or storage of information in short-term memory. According to the physics-based theory the passive aspect of conscious process merely actualizes an event that occurs in accordance with some brain-controlled rule, and this rule-selected process then develops automatically, with perhaps some occasional monitoring. Thus the theory would predict that the process of stabilization or storage in short term in memory of a certain sequence of stimuli should be able to persist undiminished while the central processor is engaged in another task.

This is what the data indicate. Pashler remarks [p.341] that "These conclusions contradict the remarkably widespread assumption that short-term memory capacity can be equated with, or used as a measure of, central resources." In the theory outlined here short-term memory is stored in patterns of brain activity, whereas consciously directed actions are associated with the active selection of a sub-ensemble of quasi-classical states. This distinction seems to account for the large amount of detailed data that bears on this question of the relationship of the stabilization or storage of information in short-term-memory to the types of tasks that require the willfully directed actions [pp. 337-341]. In marked contrast to short-term memory function, storage or retrieval of information from long-term memory, is a task that requires actions of just this sort. [pp. 347-350].

Deliberate storage in, or retrieval from, long-term memory requires willfully directed action, and hence conscious effort. These processes should, according to the theory, use part of the limited processing capacity, and hence be detrimentally affected by a competing task that makes sufficient concurrent demands on the central resources. On the other hand, "perceptual"' processing that involves conceptual categorization and identification without willful conscious selection should not be interfered with by tasks that do consume central processing capacity. These expectations are what the evidence appears to confirm: "the entirety of...front-end processing are modality specific and operate independent of the sort of single-channel central processing that limits retrieval and the control of action. This includes not only perceptual analysis but also storage in STM (short term memory) and whatever processing may feed back to change the allocation of perceptual attention itself [p. 353]."

Pashler speculates on the possibility of a neurophysiological explanation of the facts he describes, but notes that the parallel versus serial distinction between the two mechanisms leads, in the classical neurophysiological approach, to the questions of what makes these two mechanisms so different, and what the connection between them is [p.354-6, 386-7].

After considering various possible mechanisms that could cause the central bottleneck, Pashler [p.307-8] concludes that "the question of why this should be the case is quite puzzling." Thus the fact that this bottleneck and its basic properties seems to follow automatically from the same laws that explain the complex empirical evidence in the fields of classical and quantum physics means that the theory being presented here has significant explanatory power for the experimental data of cognitive psychology. Further, it coherently explains aspects of the data that have heretofore not been adequately addressed by currently applicable theoretical perspectives.

These features of the phenomena can perhaps be explained by some classical-physics-based model. But the achievement of such an explanation is hindered by the absence from classical physics of the notion of conscious choice and effort, and of the causal efficacy of conscious thoughts, and of the change-inhibiting effect of attention density. These consciousness-connected features would have to be injected ---unnaturally --- into the consciousness-free causal structure of classical theory, rather than being recognized as already existing and specified features of the causal structure of fundamental physical theory.

7. Application to neuropsychology

Quantum physics works better in neuropsychology than its classical approximation because, just as in atomic physics, it inserts knowable choices made by human agents directly into the dynamics in place of unknowable-in-principle microscopic variables. To illustrate this point we apply the quantum approach to the experiment of Ochsner et al. (2002).

Reduced to its essence this experiment consists first of a training phase in which the subject is taught how to distinguish, and respond differently to, two instructions given while viewing emotionally disturbing visual images: ATTEND (meaning passively "be aware of, but not try to alter, any feelings elicited by") or REAPPRAISE (meaning actively "reinterpret the content so that it no longer elicits a negative response"). The subjects then perform these mental actions during brain data acquisition. The visual stimuli, when passively attended to, activate limbic brain areas and when actively reappraised, activate prefrontal cerebral regions.

From the classical materialist point of view this is essentially a conditioning experiment, where, however, the "conditioning" is achieved via linguistic access to cognitive faculties. But how do the cognitive realities involving "knowing," "understanding," and "feeling" arise out of motions of the miniature planet-like objects of classical physics, which have no trace of any experiential quality? And how do the vibrations in the air that carry the instructions get converted into feelings of understanding? And how do these feelings of understanding get converted to conscious effort, the presence or absence of which determine whether the limbic or frontal regions of the brain will be activated?

Within the framework of classical physics these connections between feelings and brain activities remain huge mysteries. The materialist claim (Karl Popper called this historicist prophecy "*promissory materialism*") is that *someday* these connections will be understood. But the question is whether these connections will ever be understood in terms of a physical theory that is known to be false, and that, moreover, is false in ways that, according to contemporary physical theory, *systematically exclude the causes* of the correlations between the psychological and physiological aspects of the mind/consciousness-brain system that these neuropsychology experiments demonstrate. Or, on the other hand, will the eventual understanding of this linkage recognize and exploit the causal linkage between mental realities and brain activities that orthodox (von Neumann) contemporary physical theory specifies.

There are important similarities and also important differences between the classical and quantum explanations of the experiments of Ochsner et al. (2002). In both approaches the atomic constituents of the brain can be conceived to be collected into nerves and other biological structures, and into fluxes of ions and electrons, which can all be described reasonably well in essentially classical terms. In the classical approach the dynamics must in principle be describable in terms of the local deterministic classical laws that govern these quantities.

The quantum approach is fundamentally different. In the first place the idea that all causation is *fundamentally mechanical* is dropped as being prejudicial and unsupported either by direct evidence or by contemporary physical theory. The quantum model of the human person is essentially dualistic, with one of the two components being described in psychological language and the other being described in physical terms. The empirical/phenomenal evidence coming from subjective reports is treated as data pertaining to the psychologically described component of the person, whereas the data from objective observations, or from measurements made *upon that person,* are treated as conditions on the physically described component of the person. The apparent causal connection manifested in the experiments between these two components is then explained by the causal connections between these components *specified by the quantum laws*.

The quantum laws, insofar as they pertain to empirical data, are organized around *events* that increase the amount of information lodged in the psychologically described component of the theoretical structure. The *effects* of these psychologically identified events upon the physical state of the associated brain are specified by Processes 1 (followed by "Nature's statistical choice" of which of the discrete options specified by Process 1 will be experienced.) When no effort is applied, the temporal development of the body/brain will be roughly in accord with the principles of classical statistical mechanics, for reasons described earlier in connection with the strong *decoherence* effects. But important departures from the classical statistical predictions can be caused by conscious effort. This effort can cause to be held in place for an extended period a pattern of neural activity that constitutes a *template for action*. This delay can tend to cause the specified action to occur. In the Ochsner experiments the effort of the subject to "reappraise" *causes* the "reappraise" template to be held in place, and the holding in place of this template *causes* the suppression of the limbic response. These causal effects are consequences of the quantum rules. Thus the "subjective" and "objective" aspects of the data are tied together by quantum rules that *directly specify the causal effects of the choices made by the subject, without needing to specify how these choices came about:* the form of the quantum laws accommodates a natural dynamical breakpoint between the *cause* of willful action and its *effects*.

Quantum theory was designed to deal with cases, in which the conscious action of an agent – to perform some particular probing action - enters into the dynamics in an essential way. Within the context of the experiment by Ochsner et al. (2002), quantum theory provides, via the Process 1 mechanism, an explicit means whereby the successful effort to "rethink feelings" actually causes - by catching and actively holding in place - the prefrontal activations critical to the experimentally observed deactivation of the amygdala and orbitofrontal cortex. The resulting *intention-induced modulation* of limbic mechanisms that putatively generate the frightening aversive feelings associated with passively attending to the target stimuli is the key factor necessary for the achievement of the emotional self-regulation seen in the active cognitive reappraisal condition. Thus, within the quantum framework, the causal relationship between the mental work of mindfully reappraising and the observed brain changes presumed to be necessary for emotional self-regulation is *dynamically* accounted for. Furthermore, and crucially, it is accounted for in ways that fully allow for communicating to others the means utilized by living human experimental subjects to attain the desired outcome. The classical materialist approach to these data, as detailed earlier in this article, by no means allows for such effective communication. Analogous quantum mechanical reasoning can of course be utilized *mutatis mutandis* to explain the data of Beauregard (2001) and related studies of self-directed neuroplasticity (see Schwartz & Begley, 2002).

## 8. Conclusions

Materialist ontology draws no support from contemporary physics. The notion that all physical behavior is explainable in principle solely in terms of a local mechanical process is a holdover from physical theories of an earlier era. It was rejected by the founders of quantum mechanics, who introduced crucially into the basic dynamical equations choices that are not determined by local mechanical processes, but are attributed rather to human agents. These orthodox quantum equations, applied to human brains in the way suggested by John von Neumann, provide for a causal account of recent psycho-physical and neuropsychological data. In this account brain behavior that appears to be caused by mental effort is actually caused by mental effort: the causal efficacy of mental effort is no illusion. Our willful choices enter neither as redundant nor epiphenomenal effects, but rather as fundamental dynamical elements that have the causal efficacy that the objective data appear to assign to them, and that the subjects directly experience.

Shifting to this pragmatic approach may be as important to progress in neuroscience and psychology as it was in atomic physics.

## Acknowledgements

The work of the second-named author (HPS) was supported in part by the Director, Office of Science, Office of High Energy and Nuclear Physics, Division of High Energy Physics, of the U.S. Department of Energy under Contract DE-AC03-76SF00098. The work of the third-named author (MB) was supported in part by a chercheur-boursier from the Fonds de la Recherche en Santé du Québec (FRSQ).